\documentstyle[11pt,newpasp,twoside,epsf,graphicx]{article}
\def\edcomment#1{\iffalse\marginpar{\raggedright\sl#1\/}\else\relax\fi}
\reversemarginpar

\begin{document}
\title{A warped accretion disc in PX And ?}
\author{H.M.J. Boffin$^1$, V. Stanishev$^{2,1}$, Z. Kraicheva$^2$, \& V. Genkov$^2$}
\affil{$^1$ Royal Observatory of Belgium, 3 av. Circulaire, 1180 Brussels, 
Belgium; email: Henri.Boffin@oma.be\\
$^2$ Institute of Astronomy, Bulgarian Academy of Sciences, \\
72, Tsarighradsko Shousse Blvd., 1784 Sofia, Bulgaria}

\begin{abstract}
We have undertaken a photometric study of the SW Sex star, PX And. We clearly 
identify a negative superhump signal which might be regarded as the 
signature of a nodal precessing disc, possibly warped. PX And is also observed to possess highly
variable eclipse depth and we discuss two possible explanations.
\end{abstract}

\section{Introduction}
PX And is probably one of the most complicated SW Sex stars.
Thorstensen et al. (1991) reported shallow eclipses with highly
variable eclipse depth and repeating with a period of $\sim$0\fd1463533.
Assuming a steady-state accretion disc 
and $q\simeq0.46$,
 and fitting a mean eclipse, they obtained
$i\simeq73.8^\circ$ and $r_{\rm d}\simeq0.6R_{L_1}$.
Apart from the distinctive characteristics
of SW Sex stars (Hellier 2000),
PX And shows some other interesting peculiarities.
Patterson (1999) reported PX And to show
simultaneously ``negative" and ``positive" superhumps, and signals
with typical periods of 4--5 days.
With $q\simeq0.46$, i.e. above the canonical 0.33 value where the 3:1 resonance
still lies inside the Roche lobe, it is not clear however how PX And could
show positive superhumps.

We have undertaken a photometric study of PX And. In total  8 runs were obtained
in 2000 and 2001, with the 2.0-m telescope of Rozhen Observatory, in the Johnson $V$ filter. The exposure time used was between 20 and 40 s.
In addition, two unfiltered runs were obtained with the 1-m
telescope at Hoher List Observatory.
A detailed description of these observations, in which we clearly identify 
a negative superhump signal and its associated precession period,
are presented in 
Stanishev et al. (2002). Here, we will briefly summarize our results and then
discuss them in light of the warped precessing disc model.

\section{Results}
From our five photometric runs in October 2000, the following periods were
determined: $P^-_{\rm SH}$=0\fd142\,$\pm$0.002,
$P_1$=0\fd207\,$\pm$0.004 and $P_{\rm prec}$=4\fd8 with
semi-amplitudes of  0.086, 0.076 and 0.256 mag, respectively. 
The light curves as well as our fit to the data is shown in Fig.~1a.
Our observations also show that the star presents eclipses whose depth
is modulated with the 5-days period (Fig. 1b). The best fit to the 
depths of the eclipse is shown with a dotted line in Fig.~1b.
The mean eclipse depth is $\sim$0.56 ($\sim$0.89 mag) and the
amplitude of the variation is $\sim$0.22.

For the three runs obtained in
2001, we obtain a strong peak, corresponding to a ``negative
superhump" with $P^-_{\rm SH}\simeq$0\fd141. This value of
$P^-_{\rm SH}$ gives $P_{\rm prec}\simeq$3\fd8. The corresponding
semi-amplitudes are 0.069 and 0.21 mag. The 0\fd207 period is not
detected in 2001.

 \begin{figure}[t]
\begin{center}
   \includegraphics*[width=13cm]{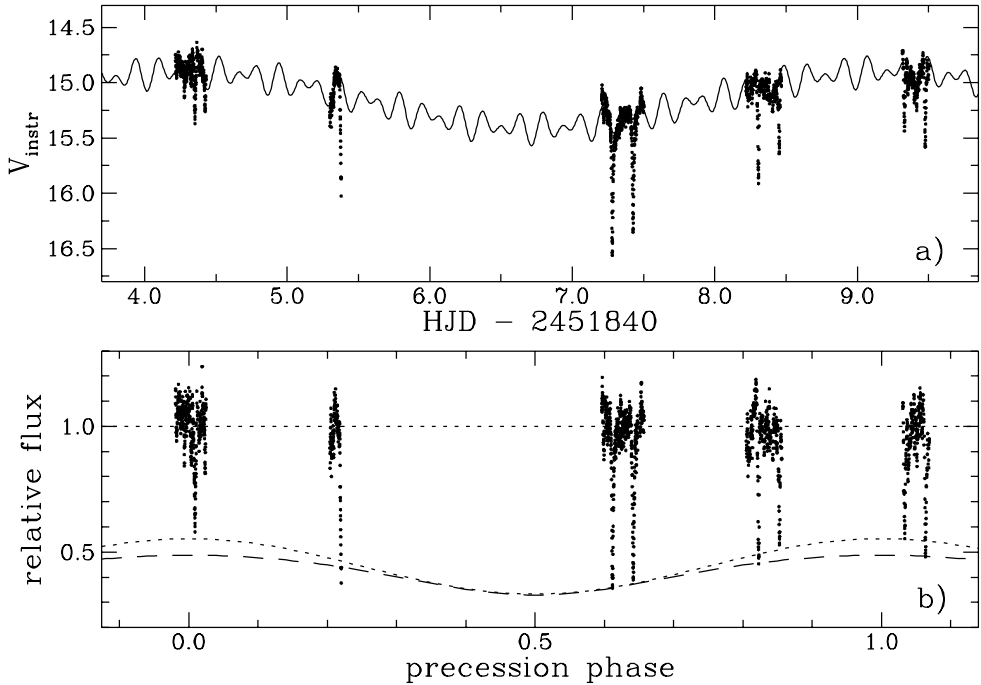}
   \caption{
{\it Upper row:} $V$ band observations of PX And and the
 best fit with the detected periods. {\it Lower row:} The same
 data, but with the best fit subtracted and displayed in flux
 scale. The fit to the eclipse depths is also shown (dotted line) as well 
 as the result of a simple model for the eclipse of a warping disc 
(dashed line; see Sect. 4.2).
   \label{eclch}
   }
\end{center}
 \end{figure}

\section{Negative superhumps}

The origin of the ``negative superhumps" is still an open
question. The most plausible model is based on the assumption of a
retrograde precession of a tilted accretion disc. However, all attempts to
simulate ``negative superhumps" failed in the sense that they were
not able to produce a significant tilt starting from a disc lying
in the orbital plane (Murray \& Armitage 1998; Wood, Montgomery, \& Simpson  2000).
The only exception we are aware of, is the recent work of Murray et al. (2002).
Once tilted, however, the accretion disc starts
precessing in retrograde direction (Larwood et al. 1996;
Wood et al. 2000).  
Taking into
account foreshortening and limb-darkening, one can estimate the
disc tilt needed to produce the observed amplitude of the 4\fd8
modulation in PX And. With a limb-darkening coefficient
$u=0.6$ the tilt angle is between 2.5$^\circ$ and 3$^\circ$,
depending on the assumed system inclination.

The mechanism generating the ``negative superhump" light itself is 
however rather uncertain. According to the simulations of Wood et
al. (2000) the two opposite parts of the disc which are
most displaced from the orbital plane are tidally heated by the
secondary. The heating is maximal when these parts point to the
secondary and this happens twice per superhump cycle. 
 Figure\,2 shows a sketch of
the system configuration in eclipse at $\phi_{\rm prec}=0.13$, that is,
when the maxima of the superhumps coincide with the eclipses.
 The expected position of the superhump light source
(SLS) according to Wood et al. (2000) and the line of nodes
are also shown. The dashed line marks the part of the disc which
lays below the orbital plane.

There is another mechanism which could generate ``negative
superhumps". Patterson
(1999) noticed that most of the SW Sex novalikes show
``negative superhumps" and this could naturally be explained by the
accretion stream overflow thought to be responsible for the SW Sex
phenomenon (Hellier \& Robinson 1994). In a system with a
precessing tilted accretion disc, the amount of gas in the
overflowing stream will vary with the ``negative superhumps"
period. Correspondingly, the intensity of the spot (shown in
Fig.\,2) formed where the overflowing stream re-impacts
the disc should be modulated and might be the ``negative" SLS.

In fact, stream overflow might even not be necessary. Indeed, if the 
disc is precessing out of the orbital plane, the "hot spot"
(i.e. the place where the L1 stream hits the accretion disc) 
will be at a different location during the precession cycle. When the disc is 
out of the plane, the hot spot will be closer to the white dwarf 
than when the disc is in the plane. Thus the amount of gravitational 
energy which will be converted partly in radiation at the hot spot will
vary with the precession phase, and this could create a modulation.
Such a model was already proposed for the intermediate polar TV Col
by Barrett, O'Donoghue, \& Warner (1988).

 \begin{figure}[tbh]
\begin{center}
\includegraphics*[width=9cm]{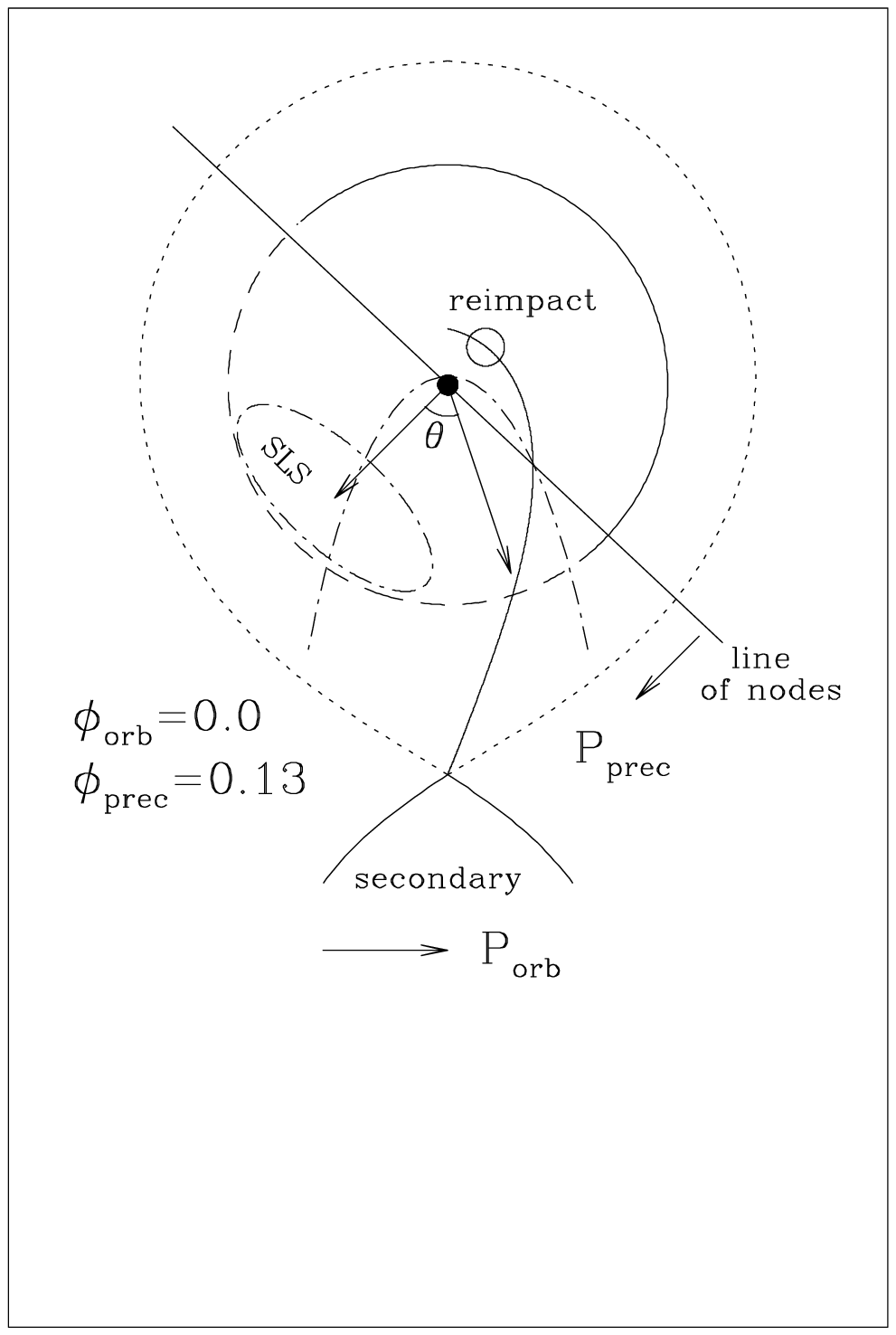}
\caption{A sketch of the system configuration at $\phi_{\rm
 prec}$=0.13 and $\phi_{\rm orb}$=0. The disc edge marked with the
dashed line lays below the orbital plane. Also shown are the
 expected position of the SLS, shadow of the secondary, accretion
 stream and re-impact zone.\vspace{0.5cm}
\label{pxmod}}
\end{center}
 \end{figure}

\section{Eclipse depth variations}

The most puzzling observation of PX And is its highly
variable eclipse depth. 
Our observations suggest that the eclipse
depth is modulated with the precession period (Fig.\,1),
indicating that this phenomenon could be
related to the disc precession. 
We will discuss two ways to modulate
the fraction of the eclipsed light with the precession period and hence
the eclipse depth.

\subsection{Veiling by superhumps light source}

If the SLS is never eclipsed, the eclipse depth will be
affected in a way analogous to
the {\it veiling} of the absorption lines in T Tau stars,
where the additional emission in the continuum reduces
the observed depth of the absorption lines. The eclipse depth
will depend on the fraction of superhump light at the moment
of eclipses, being least when superhump maxima coincide
with the eclipses.
At the moments of eclipses, the superhump flux
varies with the precession period and therefore an eclipse
depth modulation with the 5-days period will be observed.
If  $F_{\rm mod}$ is the superhump flux and $F_{\rm const}$ is the
accretion disc flux, then the equation
\begin{equation}
d_{\rm e}=\frac{d_{\rm e,0}}{1+F_{\rm mod}/F_{\rm const}}
\label{depth}
\end{equation}
gives the eclipse depth as a function of the additional
modulated light in the case when this light is not eclipsed.
Here, $d_{\rm e,0}$ is the eclipse depth if $F_{\rm mod}=0$.
At maximum the SLS in PX And contributes $\sim17\%$ to the total system
light, which means it could decrease the eclipse by only $\sim0.1$.
This is only a half of the observed modulation.

If we assume that the modulation is simply due to the fact that the 
position of the hot spot will change as a function of the precession phase,
as described at the end of Sect. 3., things become a little bit more
complicated than this simple model. Indeed, when the disc is most inclined,
then the hot spot might not be eclipsed, while it could be eclipsed 
when the disc is in the orbital plane. Moreover, the intensity of the
spot will vary depending on its location. This might well increase
the amplitude of the modulation to the observed value.

\subsection{Warped accretion disc in PX And?}

Let us assume that only part of the total light emitted by the disc
is modulated with the precession period.
Then, if the unmodulated light is totally eclipsed while the modulated light is
only partially, the eclipse depth will vary because
a different fraction of the total light emitted by the system will
be eclipsed at different precession phases. The deepest eclipses are
expected at the precession cycle minima, as observed.
The relatively shallow eclipses show that the accretion disc in
PX And is not totally eclipsed. Then, since the constant light source
must be eclipsed, its most likely location is close to the white
dwarf with the most obvious candidate being the inner hot
part of the disc. However, since the brightness modulation with the
precession cycle comes from more or less pure geometrical
considerations (precession of a tiled disc) it is not clear why
the emission of the inner parts of the disc could be constant.
One possible explanation might be that only the outer disc part
is tilted, while the inner part of the disc is not tilted. This means
that the disc is {\it warped} rather than simply tilted.
In this  case the emission from the inner disc will be constant and only
the rest will be involved in the 5-days modulation.
Generally, disc warping is not considered with respect to CVs.
Very recently, however, Murray et al. (2002) recognized the
potential role of the
magnetic field of the secondary as a source of warping in CVs
accretion discs. 
Murray et al.~have shown that the magnetic
field of the secondary can drive the disc out of the orbital plane and
warp it. As a consequence the disc starts precessing in retrograde
direction, exactly as is needed to generate "negative superhumps".

Let us consider the following simple model, which should
approximately represent the eclipse of a warped disc.
The emission from the inner part of the disc which contributes a
fraction $\alpha$ ($0<\alpha<1$) of the total light is held constant,
while the rest $1-\alpha$ is modulated so as to produce the observed amplitude
of the 5-days wave. The constant light is considered to be totally
eclipsed and only a fraction $\gamma$ ($0<\gamma<1$) of the modulated
part is eclipsed. The dashed line in Fig.\,1$b$ is calculated with
$\alpha=0.35$ and $\gamma=0.25$; this should, however, not be regarded as
a fit, but only as a demonstration of the capability of the model to
produce eclipse depth variations. A value of $\gamma=0.25$ 
corresponds approximately to what is
observed in a system with $q\simeq0.3-0.4$ and an inclination such as the
whole inner part of the disc is totally eclipsed. This simple model gives an eclipse depth
modulation of $\sim0.16$, which is closer to what is observed in PX And.
To produce the observed amplitude
of the 5-days wave the flux from the outer disc part should be
reduced by $\sim58\%$. With a system inclination
$\sim74^\circ$, this is achieved by wobbling the disc by $\pm4^\circ$, if
only the foreshortening is taken into account.

\subsection{Concluding remarks}

It seems that the two models have difficulties to explain
the observed amplitude of the eclipse depth modulation.
In spite of these difficulties we consider
our results as a step ahead. Thorstensen et al. (1991) could
produce eclipse depth variations only by changing the secondary radius by
$\sim$5\%, but as they noted this is clearly physically unreasonable.

If the first model discussed is responsible for the eclipse
depth modulation then the stream overflow is more likely to be the
mechanism generating the ``negative superhumps" in PX And.
This is because in a system with grazing eclipses the re-impact
zone is not eclipsed, while the SLS in the model of Wood et al. (2000)
should be at least partially eclipsed.

\section{Acknowledgements}
V.S. thanks the Federal Office for Scientific, Technical and Cultural
Affairs for a research fellowship in the framework of the bilateral S\&T 
collaboration with Central and Eastern Europe.
This work was partially supported by NFSR under project No. 715/97.

\end{document}